\input harvmac
\input epsf

\def\ie{{\it i.e.,}\ }
\def\eg{{\it e.g.,}\ }

\def\l{{\ell}}

\def\la{\lambda}
\def\bg{{\bar g}}
\def\cH{{\cal H}}
\def\bD{{\bar D}}
\def\bR{{\bar R}}
\def\ssc{\scriptscriptstyle}

\def\up{{}^{\ssc (d-1)}}
\def\upp{{}^{\ssc (4)}}
\def\newspace{AdS soliton} 
\def\Minim{Positive}
\def\minim{positive}


\Title
{\vbox{ \baselineskip 10pt
\rightline{hep-th/9808079 }
\rightline{NSF-ITP-98-076}
\rightline{McGill/98-13}}}
{\vbox{\centerline{The AdS/CFT Correspondence and a New} 
\centerline{\Minim\ Energy Conjecture for General Relativity}}}
\vskip10pt
\centerline{Gary T. Horowitz\foot{E-mail: gary@cosmic.physics.ucsb.edu}}
\medskip
\baselineskip=12pt
\centerline{\sl Institute for Theoretical Physics, UCSB,
 Santa Barbara, CA 93106}
\centerline{\sl Physics Department, University of California,
Santa Barbara, CA 93106}
\medskip
\centerline{\it and}
\medskip
\centerline{Robert C. Myers\foot{E-mail: rcm@hep.physics.mcgill.ca}} 
\medskip
\centerline{\sl Institute for Theoretical Physics, UCSB,
 Santa Barbara, CA 93106}
\centerline{\sl Department of Physics, McGill University,
Montr\'eal, PQ H3A 2T8}

\baselineskip=16pt
 
\vskip 2cm
\noindent

We examine the AdS/CFT correspondence when the gauge theory is considered
on a compactified space with supersymmetry breaking boundary conditions.
We find that the corresponding supergravity solution has a negative
energy, in agreement with the expected negative Casimir
energy in the field theory. Stability of the gauge theory
would imply
that this supergravity solution has minimum energy among all solutions
with the same boundary conditions.
Hence we are lead to conjecture a new positive energy theorem for asymptotically
locally Anti-de Sitter spacetimes.
We show that the candidate minimum energy 
solution is stable against all quadratic fluctuations
of the metric.

\Date{}

\def\np {Nucl. Phys. }
\def \pl {Phys. Lett. }

\def \prl {Phys. Rev. Lett. }
\def \pr  {Phys. Rev. }
\def \cqg {Class. Quantum Grav.}
\def \cmp {Commun. Math. Phys.}

\def \ann {Ann. Phys. (N.Y.)}
\gdef \jnl#1, #2, #3, 1#4#5#6{ {\it #1~}{\bf #2} (1#4#5#6) #3}

\lref\brill{D. Brill and G. Horowitz, \jnl \pl, B262, 437, 1991;
D. Brill and H. Pfister, \jnl \pl, B228, 359, 1989.}
\lref\abbott{L.F. Abbott and S. Deser, \jnl \np, B195, 76, 1982.}
\lref\wald{D. Sudarsky and R.M. Wald, \jnl \pr, D47, 5209, 1993;
R.M. Wald, gr-qc/9305022.}
\lref\witten{E. Witten, \jnl \cmp, 80, 381, 1981;
J.M. Nester, \jnl \pl, 83A, 241, 1981.}
\lref\posADS{G.W. Gibbons, C.M. Hull
and N.P. Warner, \jnl \np, B218, 173, 1983.}
\lref\garyhawk{S.W. Hawking and G.T. Horowitz, \jnl \cqg, 13, 1487, 1996.}
\lref\first{J. Maldacena, ``The Large N Limit of Superconformal
Field Theories and Supergravity,'' hep-th/9711200.}
\lref\gkp{S.S. Gubser, I.R. Klebanov and A.M. Polyakov, ``Gauge Theory
Correlators from Non-Critical String Theory,'' hep-th/9802109.}
\lref\wita{E. Witten, ``Anti-de Sitter Space and Holography,''
hep-th/9802150.}
\lref\witb{E. Witten, ``Anti-de Sitter Space, Thermal Phase Transition,
and Confinement in Gauge Theories,'' hep-th/980131.}
\lref\more{more AdS/CFT refs??}
\lref\witc{E. Witten, \jnl \np, B195, 481, 1982.}
\lref\prod{A. Taub, {\it Lett. Math. Phys.} {\bf 9} (1985) 243;
O. Moreschi and G. Sparling, {\it J. Math. Phys.} {\bf 27} (1986)
2402.}
\lref\newmet{D. Brill, J. Louko, and P. Peldan, \jnl \pr, D56, 3600, 1997
[gr-qc/9705012];  L. Vanzo, \jnl \pr,  D56, 6475, 1997 [gr-qc/9705004];
R. Mann, \jnl \np, B516, 357, 1998 [hep-th/9705223]}

\lref\yau{S.T. Yau, private communication.}
\lref\amanda{S.S. Gubser, I.R. Klebanov, A.W. Peet, \jnl
\pr, D54, 3915, 1996 [hep-th/960213].}
\lref\horror{see, for example:
G.W. Gibbons, ``Wrapping Branes in Space and Time,'' hep-th/9803206.}
\lref\perry{D.J. Gross and M.J. Perry, \jnl \np, B226, 29, 1983.}
\lref\igorstress{S.S. Gubser and I.R. Klebanov, \jnl \pl, B413, 41, 1997 
[hep-th/9708005].}
\lref\birrell{N.D. Birrell and P.C.W. Davies, {\it Quantum fields in
curved space} (Cambridge University Press, 1982).}
\lref\igorc{S.S. Gubser, I.R. Klebanov and A.A. Tseytlin,
``Coupling Constant Dependence
in the Thermodynamics of N=4 Supersymmetric Yang-Mills Theory,''
hep-th/9805156.}
\lref\blacktherm{G.W. Gibbons and S.W. Hawking, \jnl \pr, D15, 2752, 1977;
S.W. Hawking, in {\it General Relativity}, eds. S.W. Hawking and W. Israel,
(Cambridge University Press, 1979).}
\lref\btz{M. Ba\~nados, C. Teitelboim and J. Zanelli, \jnl \prl,
69, 1849, 1992 [hep-th/9204099];
M. Ba\~nados, M. Henneaux, C. Teitelboim and J. Zanelli, \jnl \pr,
D48, 1506, 1993 [gr-qc/9302012].}
\lref\garygib{
W. Boucher, G.W. Gibbons and G.T. Horowitz, \jnl \pr, D30, 2447, 1984.}
\lref\prep{R.C. Myers, in preparation.}
\lref\grosswit{D.J. Gross and E. Witten, \jnl \pr, D21, 446, 1980.}
\lref\jorma{D.R. Brill, J. Louko and P. Peldan, \jnl \pr, D56, 3600, 1997 
[gr-qc/9705012].}
\lref\glue{C. Csaki, H. Ooguri, Y. Oz and J. Terning, ``Glueball Mass Spectrum
from Supergravity,'' hep-th/9806021;
H. Ooguri, H. Robins and J. Tannenhauser, ``Glueballs and Their Kaluza-Klein
Cousins,'' hep-th/9806171;
M. Zyskin, ``A Note on the Glueball Mass Spectrum,'' hep-th/9806128;
R.. de Mello Koch, A. Jevicki, M. Mihailescu and J.P. Nunes,
``Evaluation of Glueball Masses from Supergravity,'' hep-th/9806125.}
\lref\grossooguri{D.J. Gross and H. Ooguri, ``Aspects of Large N
Gauge Theory Dynamics as seen by String Theory,'' hep-th/9805129.}
\lref\stable{P. Breitenlohner and D.Z. Freedman, \jnl \pl, B115, 197, 1982;
\jnl \ann, 144, 249, 1982; L. Mezincescu and P.K. Townsend,
\jnl \ann, 160, 406, 1985.}

\lref\lebrun{C. LeBrun, \jnl \cmp, 118, 591, 1988.}

\lref\anda{L. Andersson and M. Dahl, {\it Ann. Glob. Analysis and Geom.}
{\bf 16} (1998) 1 [dg-ga/9707017].}

\lref\blka{See \eg 
J. Bland and M. Kalka, {\it Trans. AMS} {\bf 316} (1989) 433.}

\lref\huwa{C.M. Hull and N.P. Warner, \jnl \np, B253, 675, 1985.}

\lref\bryo{J. Brown and J. York, \jnl \pr, D47, 1407, 1993.}
\newsec{Introduction}

There is growing evidence for a remarkable correspondence between
string theory in Anti-de Sitter (AdS) spacetime and a conformal field theory
(CFT)  \refs{\first,\gkp,\wita}.
In particular, type IIB superstring theory on $AdS_5\times S^5$
is believed to be completely equivalent to ${\cal N} =4$ super Yang-Mills
theory in four dimensions \first. For many applications, it suffices to
consider just the
low energy limit of the superstring theory, namely, supergravity. There is a
well defined total energy for any spacetime which asymptotically approaches
AdS \abbott, and part of the correspondence is that this energy agrees with
energy in the gauge theory. For solutions which approach AdS globally,
there are positive energy theorems which ensure that this energy 
cannot be negative \posADS, in agreement with the stability of the
gauge theory vacuum.

Witten \witb\ has suggested that one can describe ordinary (\ie
nonsupersymmetric) Yang-Mills gauge theory by compactifying one direction on a 
circle and requiring antiperiodic boundary conditions for the fermions around
the circle. In this case, the additional fermions and scalars would
acquire large masses leaving the gauge fields as the only low energy
degrees of freedom.
On the supergravity side, this proposal corresponds to considering
spacetimes which are asymptotically  AdS locally, but not globally. That is,
one spatial direction is compactified on a circle asymptotically.
If the spacetime topology is globally a simple product with an $S^1$ factor,
the standard approaches \posADS\ should still yield a positive energy
theorem (see, \eg \refs{\prod,\anda}).
However, if one considers more general
topologies, \eg for which the asymptotic
circle is contractible in the interior, those techniques will not
apply and hence it is uncertain if a positive energy
theorem will hold. It is known that in the case of asymptotically flat
spacetimes, 
it does not:
These boundary
conditions allow nontrivial zero \witc\ and negative \brill\ energy solutions.
In particular, (for a fixed size circle at
infinity) there are nonsingular
solutions to Einstein's vacuum field equations with arbitrarily negative
energy! Therefore, this sector of the theory is completely
unstable.\foot{This general result applies for any theory involving
Einstein gravity in higher dimensions, including {\it superstring theory}
\brill.  The closely related positive action conjecture is also
false for spacetimes which are only locally asymptotically Euclidean \lebrun.}

It is important to determine whether a similar instability arises for
spacetimes which are asymptotically locally AdS. From a mathematical
viewpoint, the latter
seems rather likely \yau. Negatively curved spaces tend to be less constrained
than those with positive (or zero) curvature \blka. One expects that
anything that is true for asymptotically flat spacetimes should also be true
for asymptotically AdS spacetimes. Of course, if the result {\it was}
true for the AdS case, it would have serious consequences in the context of
the AdS/CFT correspondence. The straightforward interpretation would be
that the supergravity analysis is making the rather dramatic prediction
that the nonsupersymmetric, strongly coupled gauge theory is unstable.
However, another possibility is that this result is an indication that
the correspondence fails with the nonsupersymmetric boundary conditions.
In the latter case, it would spoil the hope of
using supergravity to learn about ordinary gauge theory.

We will show that there is a static nonsingular solution (to Einstein's
equation with negative cosmological constant) with these boundary
conditions which has {\it negative} total energy. Rather than invalidate the
AdS/CFT correspondence, this particular
 solution has a natural interpretation in the gauge
theory. Since supersymmetry is broken by the antiperiodic boundary
conditions on the fermions, the gauge theory on $S^1 \times R^2$
is expected to have a negative Casimir energy. Comparing the negative energy
computed from supergravity and the Casimir energy in the weakly 
coupled gauge theory, we find close agreement. They have the same 
dependence on all parameters and disagree only by an overall factor of $3/4$.
This is similar to the factor of $3/4$ that was noticed previously in
comparisons of the entropy of the near-extremal three-branes \amanda.
We will show that in fact these two factors have the same origin.

The key question is whether the solution described above is the lowest energy
solution with these boundary conditions. If so, there must be a new 
positive energy theorem which ensures that the energy of all
solutions is
greater than or equal to this negative value, with equality only for our
particular
solution.  At first sight, this seems very unlikely, since the solution
we discuss does not have
constant curvature, supersymmetry, or any other distinguishing property
which have
previously characterized minimum energy solutions in general relativity.
Nevertheless, we will present evidence in favor of this new \minim\ energy
theorem. We will show that the solution is a local minimum of the
energy, \ie it is stable to small fluctuations. The existence of this new
 theorem can be
viewed as a highly nontrivial prediction of the AdS/CFT correspondence.
A complete proof would provide strong evidence for the correspondence.

The outline of this paper is as follows:
In the next section, we review the definition of energy for spacetimes
that are asymptotically AdS. In section 3, we present our solutions
with negative total energy and discuss their relation to the CFT.
Section 4 contains the statement of the new positive energy conjecture
and some evidence in favor of it. In section 5, we consider some
generalizations of the conjecture, and further discussion 
is given in section 6.

\newsec{Energy in Anti-de Sitter Spacetime}

The definition of total energy for spacetimes which asymptotically approach
AdS was first discussed in ref.~\abbott. In the following, we will adopt an
equivalent definition derived in \garyhawk\ (see also \bryo). The total
energy in general relativity is always defined relative to a background
solution which has a time translation symmetry.
Let the norm of the timelike Killing 
field be\foot{When there is more than one timelike Killing field,
there are additional conserved quantities. The energy is then
one component of a conserved vector (or tensor).
We will focus on one timelike component and call it the energy.}
$-N^2$. The energy depends only on $N$ and on the metric of
a spacelike surface which asymptotically approaches
the background geometry.
Starting from the action and deriving the Hamiltonian keeping track
of surface terms, one finds \garyhawk:
\eqn\energy{
E= -{1\over 8\pi G} \int N (K-K_0)  }
where the integral is over a surface near infinity,
 $K$ is the trace of the extrinsic curvature of this surface,
and $K_0$ is the trace of the extrinsic curvature of a surface with the same
intrinsic geometry in the background or reference spacetime.\foot{To leading
order, $N$ will be the same for both the given metric and the background
space. Higher order differences between $N$ and $N_0$
will not affect the result for the energy \garyhawk.}
This definition is very general and works for both asymptotically flat
and asymptotically AdS spacetimes.

Let us illustrate this definition
 with a few examples.  Consider  the Schwarzschild-AdS
solution in  four dimensions:
\eqn\sads{
ds^2 = - \left({r^2\over \l^2} + 1 -{r_0\over r}\right) dt^2
 + \left({r^2\over \l^2} + 1 -{r_0\over r}\right)^{-1} dr^2
  + r^2 d\Omega_2}
where $\l$ is related to the negative cosmological constant by
$\l^2=-3/\Lambda$.
Consider a spatial slice  of constant $t$ in this space.
At fixed $r$, one has a round two-sphere with area
$A =4\pi r^2 $. The integral of the
trace of the extrinsic curvature  of this sphere is easily computed as
\eqn\extcur{
\int K = n^\mu \partial_\mu A
=  \left({r^2\over \l^2} + 1 -{r_0\over r}\right)^{1/2} 8 \pi r  }
where $n^\mu$ is the unit radial vector normal to the sphere.
The background or reference spacetime is just anti de Sitter space,
i.e., \sads\ with $r_0=0$. At fixed $t$, the boundary surface in
the background with the same intrinsic geometry as above is again
a two-sphere at the same value of the 
radial coordinate $r$. Thus $\int K_0$ is simply given by \extcur\ with $r_0=0$.
In either case, $N$ is constant on the sphere, and asymptotically
approaches $N \simeq r/\l$.
Substituting these expressions into \energy\ yields $ E = r_0/2G_4 $ as 
expected (where $G_4$ is Newton's constant in four-dimensions).

This calculation is easily extended to arbitrary dimensions with the
black hole metric 
\eqn\sadsall{
ds^2 = - \left[{r^2\over \l^2} + 1 -\left({r_0\over r}\right)^{p-1}\right]
dt^2 + \left[{r^2\over \l^2} + 1 -\left({r_0\over r}\right)^{p-1}\right]^{-1}
{ dr^2}+ r^2 d\Omega_{p}}
where $d\Omega_{p}$ is the metric on a unit $p$-sphere, and
$\l^2=-p(p+1)/2\Lambda$. Also note that $p\ge2$ for the above metric.
The final result for the energy is
\eqn\energee{
E={p\,\Omega_{p}\over 16\pi G_{p+2}}r_0^{p-1} }
where $\Omega_{p}=2\pi^{p+1\over2}/\Gamma\left({p+1\over2}\right)$ is the
area of a unit $p$-sphere, and $G_{p+2}$ is the ($p$+2)-dimensional
Newton's constant.

Next consider the following asymptotically AdS metrics:
\eqn\neext{
ds^2 ={r^2\over \l^2}\left[-\left (1 -{r_0^{p+1}\over r^{p+1}}\right)
 dt^2 + (dx^i)^2 \right] +
 \left(1 -{r_0^{p+1}\over  r^{p+1}}\right)^{-1} {\l^2 \over r^2}dr^2 }
where $i=1,\cdots, p$. 
For certain values of $p$, these metrics arise in the near horizon geometry of 
$p$-branes (see, \eg \first). With $r_0=0$, these metrics correspond to
AdS space in horospheric coordinates \horror.
Once again we consider a surface of constant $t$.
If we introduce $V_p$ as the coordinate volume
of the surfaces
parameterized by $x^i$, then the area of a surface at fixed large $r$
is simply $A = r^p V_p/\l^p$. Computing the energy as before yields
\eqn\pbrane{
E_p =   {p\, V_{p}\over 16\pi G_{p+2} \l^{p+2}} r_0^{p+1}\ .}
 $E_p/V_p$
corresponds to the energy density of the  field theory in the
CFT/AdS correspondence.

There is a slight subtlety in computing the mass of the above metrics
\neext. If the
directions along the brane $x^i$ are not identified (\ie are noncompact),
 then the constant $r_0$
can be changed by rescaling the coordinates $t, r, x^i$
in an appropriate way. Hence the energy 
\pbrane\ is
not well defined. This is not surprising, since the energy is conjugate
to asymptotic time translations, and so if one rescales the time, the energy
should change.\foot{In the full asymptotically flat $p$-brane
solution, this is not a problem, since the scale for $t$ is picked out by the 
requirement that $\partial_t$ be a unit time translation at  infinity.
It is this time which corresponds to time translation in the gauge theory.
In the previous metrics \sadsall, the scale of $r$ is fixed
by requiring the spheres of constant radius to have an area given by
$r^p \Omega_p$.}
In the following, we will be interested in the case where at least one of the
spatial directions
is compactified.
 If we fix the periodicity of the circle (corresponding to
fixing the size of the circle in the gauge theory) then $r_0$ cannot be
rescaled. However, when some of the $x^i$'s are
compactified, the background spacetime with
$r_0 = 0$ has a conical singularity at $r=0$. We will not worry about
this singularity, since it is likely that string theory resolves it
without changing the asymptotic form of the metric, which is all that is needed
to compute the energy. More importantly, the lower energy solution we
describe in the next section is completely nonsingular.
	
\newsec{Negative Energy Solutions}

We begin by reviewing the negative energy solutions in the asymptotically
flat context \brill.
It is easy to describe the
initial data for these negative energy solutions.
For five dimensional solutions, the initial data consists of a four-dimensional
Riemannian manifold which asymptotically approaches the flat metric on $S^1
\times R^3$. 
Of course, within general relativity, this initial data must satisfy
a number of constraint equations. However,
if we set the conjugate momentum to zero, these constraints
reduce to the condition that the scalar curvature
vanish. As initial data, we 
consider the euclidean Reissner-Nordstr\"om metric
\eqn\rn{
ds^2 = \left( 1 - {2m\over r} + {q^2\over r^2} \right) d\tau^2
  +\left( 1 - {2m\over r} + {q^2\over r^2}\right)^{-1} dr^2
  + r^2 d\Omega_2 }
To avoid a conical singularity at $ r = r_+ \equiv m +\sqrt{m^2 -q^2}$,
we must periodically identify $\tau$ and so \rn\ has the desired asymptotic
geometry. It also satisfies the constraint, because the Einstein tensor is
proportional to the Maxwell stress tensor, which is trace-free in
four dimensions. We now
analytically continue the parameter $q\rightarrow iq$.
(Since we are interested only in
the metric \rn\ and do not include a Maxwell field, we do not have to worry
about the latter becoming 
complex.) It is now clear that we can take the mass parameter $m <0$ 
without the metric becoming singular. Since the size of the circle at
infinity is just the period of $\tau$ which depends on both $m$ and $q$,
one can keep this fixed as $m$ becomes arbitrarily negative. In fact, one
finds, for a fixed period, that the curvature remains bounded as the mass
becomes increasingly negative. Therefore one may conclude that
nonsupersymmetric compactifications in asymptotically 
flat spacetimes are unstable. This analysis is quite general and may
be applied to any theory involving Einstein gravity in higher dimensions,
including superstring theory \brill.

We now want to know if an analogous result holds for spacetimes which are
asymptotically AdS. The first thing to try is the obvious generalization
of the above procedure using the
euclidean AdS Reissner-Nordstr\"om metric. When the charge in the latter
is analytically continued, the metric becomes
\eqn\rnads{
ds^2 = \left({r^2\over \l^2}+ 1 - {r_1\over r} -
 {r_0^2\over r^2} \right) d\tau^2
 +  \left({r^2\over \l^2}+ 1 - {r_1\over r} - {r_0^2\over r^2}\right)^{-1} dr^2
  + r^2 d\Omega_2 }
As before, this satisfies the vacuum constraints (now with negative cosmological
constant), if the momenta are set equal to zero. However, there is an
important difference with the asymptotically flat case.
In \rnads,   the proper length of the circles
parameterized by $\tau$ grows with $r$. This means that the area of the
surface at infinity grows like $r^3$ just like the uncompactified
five-dimensional AdS. As a result, the mass is determined by the $r_0^2/r^2$
terms in the metric, rather than the $r_1/r$ term. 
The appropriate physical boundary conditions -- see the discussion in section
5 -- require that $r_1=0$.
Thus, this construction only yields the following one parameter
 family of finite energy initial data:
\eqn\rnadsn{
ds^2 = \left({r^2\over \l^2}+ 1  -
 {r_0^2\over r^2} \right) d\tau^2
 +  \left({r^2\over \l^2}+ 1 - {r_0^2\over r^2}\right)^{-1} dr^2
  + r^2 d\Omega_2 }
Note that one cannot change the sign of $r_0^2$ without introducing a
naked singularity at $r=0$. With the above sign,
the radial coordinate is restricted to $r\ge r_+$ where $r_+$ is the largest
root of $r_+^4 + \l^2(r_+^2 - r_0^2)  =0$. To avoid a conical singularity
at $r= r_+$, $\tau$ is identified with period
\eqn\period{
\beta = {2\pi \l^2 r_+\over 2r_+^2 + \l^2}\ .}
The metric \rnadsn\ can also be obtained by analytically
continuing  the five-dimensional Schwarzschild AdS solution and restricting 
to the equatorial plane of the three-spheres.
One might thus expect
that its mass  would be positive.
However, we now show that it is negative!

The area of a surface at large $r$ is
\eqn\morarea{
A =  4\pi r^2 \beta\left({r^2\over \l^2}+ 1 - {r_0^2\over r^2}\right)^{1/2}}
The integral of the extrinsic curvature becomes
\eqn\extcurvs{
\int K = 4\pi \beta\left[ {3r^3\over \l^2} + 2r -{r_0^2\over r} \right]}
The background metric is simply \rnadsn\ with $r_0  =0$, which is
four-dimensional hyperbolic space with periodic identifications\foot{Even
though the spacetime resulting from this initial data is locally AdS, it is
not globally static. So extra restrictions are needed to ensure energy
conservation. This will not be the case for our main example discussed below.}.
In this reference space, we need to choose a boundary surface with
the same intrinsic geometry as the $S^2\times S^1$ at fixed $r$ above.
For the $S^2$ geometry to agree, the radial coordinate of the surface
in the background must be the same
as in the original spacetime. For the proper distances along the $S^1$ factors
to agree, the periodicity of $\tau$ in the background $\beta_0$ is related to 
$\beta$ by
\eqn\relbeta{
 \left({r^2\over \l^2}+ 1 - {r_0^2\over r^2}\right)^{1/2} \beta\  = \
 \left({r^2\over \l^2}+ 1 \right)^{1/2} \beta_0 }
The integral of the extrinsic curvature in the background is simply
\eqn\extbac{
\int K_0 = 4\pi \beta_0 \left[ {3r^3\over \l^2} + 2r\right] }
Using \relbeta, $N\simeq r/\l$, and the definition of the energy \energy,
one finds that
\eqn\maschw{
E = -{\beta r_0^2\over 4 G_5 \l} }
{}From this, one clearly sees that the energy is negative, but it is difficult
to see the dependence on the size of the circle $\beta$ since $r_0$
is implicitly related to $\beta$ through \period\ and the definition of $r_+$.
For $r_+\gg \l$, $ r_0^2 \simeq r_+^4/\l^2 \simeq \pi^4\l^6/\beta^4$,
and so one obtains\foot{For a given $\beta$, there is another solution to
\period\ with a smaller value of $r_+$, but it corresponds
to a configuration for which the energy which is less negative.}
\eqn\finalexp{
E\simeq- {\pi^4 \l^5 \over 4 G_5 \beta^3}\ . }

Through the AdS/CFT correspondence,
this analysis should be related to a gauge theory on $S^2 \times S^1$ where
$S^2$ has radius $\l$, $S^1$ has period $\beta$, and 
supersymmetry breaking boundary conditions are imposed along the $S^1$.
For small
$\beta$, this is expected to have a negative Casimir energy density
proportional to $\beta^{-4}$ (at least at weak coupling). Hence
the total energy would be negative and proportional to $\beta^{-3}$,
in agreement with the above supergravity calculation.

The preceding calculations can be extended to arbitrary dimensions
with the initial data metric
\eqn\highads{
ds^2 = \left[{r^2\over \l^2}+ 1  -
 \left({r_0\over r}\right)^{p-1} \right] d\tau^2
 +  \left[{r^2\over \l^2}+ 1  - \left({r_0\over r}\right)^{p-1}\right]^{-1}
 dr^2+ r^2 d\Omega_{p-1} }
which satisfies the constraint equation ${}^{(p+1)}R=-p(p+1)/\l^2$.
Again, the geometry smoothly closes off at $r=r_+$, which is now the
largest root of $r_+^{p+1}+\l^2(r_+^{p-1}-r_0^{p-1})=0$, provided that
$\tau$ is identified with period
\eqn\newperiod{
\beta = {4\pi \l^2 r_+\over (p+1)r_+^2 + (p-1)\l^2}\ .}
The final result for the energy is
\eqn\newenergy{
E=-{\Omega_{p-1}\beta r_0^{p-1}\over 16\pi G_{p+2}\l}
\simeq - {\Omega_{p-1}\over16\pi G_{p+2}}\left({4\pi\over p+1}\right)^{p+1}
{\l^{2p-1}\over \beta^p} }
where the final formula again holds for large $r_+$. Where applicable,
these results should be related to a  quantum
field theory on $S^{p-1}\times
S^1$. Again, a negative energy proportional to $\beta^{-p}$ can be expected
to arise through the Casimir effect in the field theory.

To make a more precise comparison of the energy in AdS and the gauge theory,
it would be useful to have an example of a solution with negative energy
that asymptotically had topology $R^{p-1} \times S^1$. This is easily obtained
by a double 
analytic continuation of the near-extremal $p$-brane solution \neext.
That is, we analytically continue this metric with both $t\rightarrow i\tau$
and $x^p\rightarrow i t$. In the following, we will
refer to this spacetime as the {\it AdS soliton}.
The metric becomes
\eqn\aneext{
ds^2 ={r^2\over \l^2}\left[\left (1 -{r_0^{p+1}\over r^{p+1}}\right)
 d\tau^2 + (dx^i)^2 -dt^2  \right] +
 \left(1 -{r_0^{p+1}\over  r^{p+1}}\right)^{-1} {\l^2\over r^2}dr^2 }
where there are now $p-1$ $x^i$'s. Again, the coordinate $r$ is restricted to
$r\ge r_0$ and $\tau$ must be identified with period
$\beta = 4\pi l^2/ (p+1)r_0$ to avoid a conical singularity
at $r=r_0$. Note that this spacetime metric is globally static and
completely nonsingular.
For fixed $t$ and $r$, the area of a surface
is $A = \left (1 -{r_0^{p+1}\over r^{p+1}}\right)^{1/2} r^p \beta V_{p-1}/\l^p$,
where $V_{p-1}$ denotes the volume of the transverse $p-1$ $x^i$'s.
The appropriate background is \aneext\ with $r_0 =0$, which
corresponds to AdS space with periodic identifications.
It is easy to see
that for the background, $\int K_0 = (p/\l) A$. Following the above procedure,
the energy is again found to be negative
\eqn\massane{
E = -  { r_0^{p+1} \beta V_{p-1}\over 16 \pi G_{p+2} \l^{p+2}} }
Using the above relation between $r_0$ and $\beta$, this result
can be rewritten as 
\eqn\masane{
E = -  { V_{p-1}\l^p\over 16 \pi G_{p+2} \beta^p }
\left(4\pi \over p+1\right)^{p+1}\ . }

For certain values of $p$, we can express this in terms of the
string theory coupling $g$ and Ramond-Ramond charge $N$. 
For example, for the three-brane, $p=3$,
we have
$\l^4 = 4\pi gN$, and $G_5$ is the ten-dimensional Newton's constant,  
$G_{10} = 8 \pi^6 g^2$,
divided by the volume of a five sphere of radius $\l$, $A_5 = 
\pi^3 \l^5$. We thus obtain an energy density
\eqn\finaleng{
\rho_{\rm sugra} ={E\over V_2\beta} = -{\pi^2 \over 8}
{N^2\over\beta^4}\ . }

We wish to  compare this with the ground state energy of the 
gauge theory
on $S^1 \times R^2$, where the length of the $S^1$ is $\beta$. 
 This can only be calculated directly at weak gauge 
coupling, where to leading order, it reduces
to the problem of determining the Casimir energy of the free field
theory. The field theory is $N=4$ super-Yang Mills, which
contains an SU(N) gauge field, six scalars in the adjoint representation,
and their superpartner fermions. In the present case, the latter fermions
are antiperiodic on the $S^1$. The stress-energy tensor for this theory
may be found in \igorstress. The leading order
Casimir energy may be calculated by point-splitting the fields in the
energy density (\ie $T_{tt}$) with the appropriate
free-field Green's function and then removing the vacuum divergence before
taking the limit of coincident fields \birrell.
The final result is
\eqn\gauge{\rho_{gauge} = -{\pi^2\over 6}{N^2\over \beta^4}\ . } 
Thus we find that the negative energy density of the supergravity solution is
precisely $3/4$ of the Casimir energy of the weakly coupled gauge theory!
This is very reminiscent of the earlier results showing that the entropy of the
near-extremal three-brane is precisely $3/4$ the entropy of the weakly
coupled gauge theory at the same temperature \amanda.

In retrospect, it is not surprising that the ground state energies differ
by exactly the same factor as the thermal entropies, as both results can be
derived as different interpretations of a common euclidean calculation. 
On the field theory side, consider the euclidean functional integral for the
(weakly coupled) Yang-Mills theory on $S^1\times R^3$ where the circle
has period $\beta$ and antiperiodic boundary conditions for the fermions.
 One natural interpretation is as a
thermal field theory calculation at temperature $\beta^{-1}$, and the
partition function yields the free energy as $\beta F_{YM} =-\log Z_{YM}$.
Alternatively, one may interpret one of the noncompact directions as
euclidean time $t_E$. In this case the same partition function, 
evaluated between two surfaces
separated by a large difference of euclidean time $\Delta t_E$, yields the
ground state energy as $\Delta t_E\, E_{YM}=-\log Z_{YM}$. On the supergravity
side, consider the euclidean instanton obtained by analytically continuing
the \newspace\ metric \aneext\ with $t \rightarrow i\ t_E$. This instanton
is, of course, identical to that obtained from the black hole metric
\neext\ with $t\rightarrow i \tau$ and identifying $\tau$ with the
appropriate period $\beta$. (We also trivially rename $x^p=t_E$.)
In the latter context, the instanton describes a thermal equilibrium
of the black hole \blacktherm\ at temperature $\beta^{-1}$, and
the euclidean action is interpreted as giving the black hole contribution
to the free energy as $\beta F_{BH} =I$. 
On the other hand, in the context of the \newspace,
the same euclidean action,  is
simply related to the total energy \masane\ via $\Delta t_E\, E=I$.
Now from the analysis of three-branes \amanda\ ($p=3$), 
it follows that when the temperatures of the black
hole and Yang-Mills theory are equated, their free energies are related by: 
\eqn\freeng{
F_{BH}  =  {3\over 4}  F_{YM}  }
and hence
\eqn\eucaction{
I = {3\over 4} (-\log Z_{YM})\ . }
Hence from the preceding discussion, it also follows that we must find
the same factor in relating the total energies, $E= 3/4\, E_{YM}$, and
the energy densities above, as well.

The factor of $3/4$ discrepancy between the two calculations
does not contradict the AdS/CFT correspondence.
Rather the supergravity
result \finaleng\ corresponds to the energy density of the gauge
theory in a regime of strong coupling. To
extrapolate the AdS results to weak coupling, one must include all of the
higher order (in the string scale) corrections to the geometry induced
by the Type IIB string theory. The leading order
correction to 
\aneext\ has recently been computed in \igorc. The
net effect is that the euclidean action became more negative.
Thus from the preceding discussion, as expected, the energy 
and energy density of the
corresponding \newspace\ becomes slightly more negative, improving the
agreement with the weak coupling results \gauge.

In order to construct these negative energy solutions \aneext,
we need $p\ge 1$ so there will be one spatial direction in \neext\ to
analytically continue. The case $p=1$,  
corresponding to three spacetime dimensions, is special.
All solutions have 
constant curvature and hence are locally AdS. The metric \neext\
with $p=1$ is the (nonrotating) BTZ black hole \btz. We showed above that
if you take the positive mass black hole and double analytically continue,
then you get a solution with less mass than the zero mass black hole.
However, it is known that
 three-dimensional AdS spacetime itself has less mass than the 
$M=0$ black hole \btz. 
In 
fact, as we will now show, the double analytically continued black hole is
precisely AdS globally, with no extra identifications. 
We start with the black hole metric \neext\ with $p=1$
\eqn\tdbh{
ds^2 = -{r^2 - r_0^2\over\l^2} dt^2 + {\l^2\over r^2 - r_0^2} dr^2 
+ {r^2\over\l^2} dx^2  }
By rescaling $t,r, x$, we can set $r_0 = \l$. Now analytically continue
in $t$ and $x$ as before to get
\eqn\tdancon{
ds^2 = -{r^2\over\l^2} dt^2 + \left({r^2\over\l^2} - 1\right)^{-1} dr^2
+ \left({r^2\over\l^2} - 1\right) d\tau^2  }
Finally set $\rho^2 = r^2-\l^2$  and $\phi=\tau/\l$ to put this metric
into the standard AdS form
\eqn\tdads{
ds^2 = -\left({\rho^2\over\l^2} +1\right) dt^2 + \left({\rho^2\over\l^2} +1
\right)^{-1} d\rho^2 + \rho^2 d\phi^2 }
Note that after the analytic continuation, $\tau$ should be periodically
identified to avoid a conical singularity. In \tdads, this is simply
the statement that $\phi$ is an angle with the standard periodicity
of $2\pi$. 

\newsec{A New Positive Energy Theorem?}

\subsec{The conjectures}

The above qualitative and quantitative agreements between AdS energy
and Casimir energy in the CFT seem to support the AdS/CFT
correspondence in the  nonsupersymmetric case. A crucial
question though is whether the AdS soliton \aneext\ is the lowest energy
solution with the given boundary conditions. The aforementioned agreement
would be put in peril by the existence of metrics
with even lower energies.

For definiteness, let us focus on the $p=3$ case in the following.
This corresponds to the near horizon geometry of Dirichlet three-branes,
for which the AdS/CFT correspondence is understood in most detail.
The \newspace\ metric \aneext\  then becomes
\eqn\threebrane{
ds_3^2={r^2\over \l^2}\left[\left(1-{r_0^4\over r^4}\right) d\tau^2 
+(dx^1)^2+(dx^2)^2-dt^2\right]+
\left(1-{r_0^4\over r^4}\right)^{-1}{\l^2\over r^2}{dr^2} }
where $r\ge r_0$ and  $\tau$ has period 
$\beta = \pi l^2/r_0$.
{\it For the remainder of this section, we will use \threebrane\ as our
reference metric and measure
energy relative to it.} 
We will consider metrics which asymptotically approach \threebrane\ in the
sense that for large $r$ 
\eqn\boundary{g_{\mu\nu} = \bar g_{\mu\nu} + h_{\mu\nu} \ ,} 
$$h_{\alpha\gamma} = O(r^{-2}), \quad h_{\alpha r} = O(r^{-4}), 
\quad h_{rr} = O(r^{-6}){\rm \ \ with\ } \alpha,\gamma\ne r.$$
Derivatives of $h_{\mu\nu}$ are required to fall off one power faster.
In \boundary, $\bar g_{\mu\nu}$ denotes the AdS soliton \threebrane\
(or metrics obtained from it as indicated below). Note that even though
these boundary conditions allow metrics with different constants $r_0$
asymptotically, the periodicity of $\tau$ is fixed.

The AdS/CFT correspondence together with the expected stability of the
nonsupersymmetric gauge theory, suggests that the energy of any solution
with these boundary conditions should be positive relative to \threebrane.
Hence we are lead to formulate a new positive energy
conjecture. Below we present three different forms of this conjecture, starting
with the most general and becoming more specialized. The simpler conjectures
may be easier to prove, but would still be of great interest.
\medskip
{\bf Conjecture 1:}  Consider all solutions to ten-dimensional
 IIB supergravity satisfying \boundary\ (with
 $\bar g_{\mu\nu}$ denoting the product of \threebrane\
 with a five sphere of radius $\l$).
Then $E\ge 0$, with equality if and only if $g_{\mu\nu} = \bar g_{\mu\nu}$.
\medskip
The self-dual five form must be nonzero to satisfy the asymptotic boundary
conditions. 
If we make the reasonable assumptions that the other supergravity fields
will only increase the energy, and spacetimes which are not direct products
with $S^5$ will also have higher energy, then the
above conjecture can be reduced 
from ten dimensions to five as follows.

\medskip
 {\bf Conjecture 2:}  Consider all solutions to Einstein's 
equation in five dimensions with cosmological constant $\Lambda = -6/\l^2$
 satisfying \boundary\
(with $\bar g_{\mu\nu}$ denoting the metric \threebrane).
Then $E\ge 0$, with equality if and only if $g_{\mu\nu} = \bar g_{\mu\nu}$.
\medskip

In the above conjectures, the solutions are required to have at least
one nonsingular spacelike surface, since otherwise one could easily
construct counterexamples
with naked singularities. 
If we assume that there is a surface with zero extrinsic
curvature (\ie a moment of time symmetry) then the constraint equations
reduce to the statement that the scalar curvature is constant. We thus
obtain:

\medskip
{\bf Conjecture 3:} Given a nonsingular
Riemannian four manifold with $R = -12/\l^2$
satisfying \boundary\ (with $\bar g_{\mu\nu}$ denoting the metric 
on a $t=$ constant surface in \threebrane), then $E\ge 0$
 with equality if and only if  $g_{\mu\nu} = \bar g_{\mu\nu}$.

\medskip

As we mentioned in the introduction,
at first sight these conjectures seem unlikely to be true. The solution
\threebrane\ does not have constant curvature, supersymmetry, or any other
special property usually associated with minimum energy solutions
in general relativity.
It is possible that the above conjectures fail,
but there is another solution of minimum energy. However, this is unlikely,
since one expects the minimum energy solution to be static
and translationally invariant around the circle. One could then
double analytically continue this metric to produce a new black hole solution.
The ``black hole uniqueness theorems"
(which have not been proven for
this case, but still are believed to be true) would then imply that
this solution must be identical to \neext\ with $p=3$, which
corresponds to the analytic continuation of \threebrane. Previous experience
would suggest that there are time dependent solutions of arbitrarily
negative energy --- see, \eg \brill.

Nevertheless, in this section we present some evidence that
 the above conjectures are indeed
true. First we note that under perturbations of the metric \threebrane,
the energy is unchanged to first order. This result in fact applies
for any  metric that is globally static \wald, 
 and can be seen as follows.
The gravitational Hamiltonian is a function of the spatial metric and
conjugate momentum, and takes the form
$H(g_{ij},\pi^{ij}) = \int N^\mu C_\mu  + E  $. 
where $N^\mu$ is the lapse-shift vector and $C_\mu$ are the constraints.
Suppose we start with a static solution
and choose $N^\mu$
to generate evolution along the time translation symmetry. Consider the
variation of $H$ with respect to $g_{ij}$. On the one hand, this is
$\partial_t \pi^{ij}$ which vanishes since the background is static. On the other
hand, the variation of the constraint will vanish whenever
the perturbation solves
the linearized constraints. Hence the variation of the energy must also vanish.
Since $E$ is independent of the conjugate momenta, this
is sufficient to establish that the energy is an extremum.

\subsec{Perturbative stability}

While we have established
 that the mass of the \newspace\ is an extremum, we would
like to show that it is actually a global minimum. Unfortunately,
given the (nonsupersymmetric)
spin structure on the asymptotic geometry, we can not apply the spinor
techniques of \witten\ and \posADS\ to argue that this is the case.
Instead we must
be satisfied with showing that the \newspace\ gives a local minimum of
the AdS energy functional. Our perturbative approach here
follows that of \abbott, where the stability
of the AdS spacetime itself was considered. We refer the interested
reader there for a detailed discussion of the technique.
 Their general analysis is based on the
construction of conserved charges for background
solutions with Killing symmetries, in particular a time translation symmetry,
and hence may be applied to the \newspace. 

One begins by dividing the metric (globally) in a manner similar to \boundary
\eqn\divide{
g_{\mu\nu}=\bg_{\mu\nu}+h_{\mu\nu} }
where $\bg_{\mu\nu}$ is the AdS soliton 
and $h_{\mu\nu}$ represents a deviation satisfying the above boundary 
conditions. For the moment we will work with the general $d$-dimensional
case, and later specialize to $d=5$.
In order that $g_{\mu\nu}$ is still a
solution, $h_{\mu\nu}$ must satisfy an equation which may be represented
as a linearized Einstein equation with a nonlinear source term. Written
in this form, the terms nonlinear in $h_{\mu\nu}$ may be taken to define
the energy-momentum density of the gravitational field, $T^{\mu\nu}$.
By virtue of the field equations, this density is covariantly conserved
in the background metric, \ie ${\bar \nabla}_\mu T^{\mu\nu}=0$.
Now given a Killing vector $\xi^\mu$ of the background solution, one finds
then that $T^\mu{}_\nu\xi^\nu$ is a covariantly conserved current and
hence 
\eqn\conserve{
E(\xi)={1\over8\pi G}\int d^{d-1}x\ \sqrt {\bar g}\, T^0{}_\nu\,\xi^\nu }
is a conserved charge. If $\xi^\mu$ is a timelike vector, this
quantity defines the Killing energy, \ie the mass of the new metric \divide\ 
with respect to the background solution. Further, Abbott and Deser
show that the integrand of eq. \conserve\ is a total divergence, and
so $E(\xi)$ may be written as a flux integral over a ($d-2$)-dimensional
surface at infinity. The details of this calculation are not important
here, however, we note that with this flux integral form one may show that 
this Killing energy \conserve\ agrees with our previous definition of the energy
\energy \garyhawk.

Instead we wish to construct $E(\xi)$ (or rather the energy density
$\cH$) directly to second order in the
fluctuations $h_{\mu\nu}$. Abbott and Deser \abbott\ turn to the framework
of canonical gravity for this purpose. One may view their construction
as evaluating the second order term in the field equations by making a
third order expansion of the action with a judicious choice of variables.
The canonical variables provide a judicious choice for
several reasons. First, since one wishes to focus on $T^0{}_\mu$, the third
order expansion need only be carried out for terms linear in
the lapse or shift, and quadratic in the spatial metric and conjugate
momenta. Second, for the present background
solutions of interest, we will be evaluating $\cH$ on a
time symmetric slice, and so the background momentum variables vanish.
Finally, the background shift vector also vanishes in the solutions considered
here. The net result
is that one must calculate
\eqn\ham{
\cH={{\bar N}\over\sqrt{\bar g}}\left[g_{ik}g_{jl}\pi^{ij}\pi^{kl}
-{1\over d-2}\pi^2-g(\up R-2\Lambda)\right]}
with the quantity in square brackets evaluated to second order in the
deviations of the spatial metric $h_{ij}$ and the conjugate momentum
deviations $p^{ij}$. Here $\up R$ is the intrinsic curvature scalar
of the initial data surface. For convenience, the following gauge
conditions are imposed on the fluctuation fields \abbott
\eqn\gauge{
p^i{}_i=0=\bD^ih_{ij} }
where $\bD^i$ is the ($d-1$)-dimensional covariant derivative on the
initial data surface with the background metric. The deviations are also
required to satisfied the constraint equations to linear order, which
imposes 
\eqn\constrain{
h^i{}_i=0=\bD_ip^{ij} }
One can see that together \gauge\ and \constrain\ ensure that the
fluctuations are transverse and traceless with respect to the background
metric. Evaluating \ham\ subject to these constraints, one arrives
at the following expression\foot{One must integrate by parts to arrive
at this expression, however, the above boundary conditions \boundary\ will
ensure the vanishing of any boundary contributions.}
\eqn\hamb{
\cH={{\bar N}}\left[{1\over\sqrt{\bar g}}p^{ij}p_{ij}+\sqrt{\bar g}\left(
{1\over4}(\bD_kh_{ij})^2+{1\over2}\up{\bar R}^{ijkl}h_{il}h_{jk}-{1\over2}
\up{\bar R}^{ij}h_{ik}h_j{}^k\right)\right]\ .}

Here it is immediately apparent that the momenta  make a manifestly positive
contribution to the energy density. Hence if we are interested in lowering
the energy of the background solution we should set $p^{ij}=0$ and focus
on the spatial metric fluctuations. For the latter, there is a gradient
energy density, which is also positive, and a potential energy
density, which a priori has no definite sign. If we consider the background
to be anti-de Sitter space, for which
\eqn\adscurv{
\up{\bar R}_{ijkl}=-{1\over\l^2}(\bg_{ik}\bg_{jl}-\bg_{il}\bg_{jk})}
the potential becomes
\eqn\adspot{
U=+{d-3\over2\l^2}h_{ij}h^{ij}}
where $\l^2=-(d-2)(d-1)/2\Lambda$.
Hence in AdS space, the potential energy and hence the total energy contribution
of the spatial metric fluctuations is also manifestly positive, and we may
conclude that AdS space is perturbatively stable. Of course, spinor techniques
\witten\ allow one to show that AdS is in fact the absolute minimum energy
state within that sector of the theory, \ie for solutions
which admit asymptotically constant
spinors. The \newspace\ is not included in this
sector, as the spin structure on the asymptotic boundary differs.
Further the latter metric does not have a Riemann tensor with the
maximally symmetric form of eq. \adscurv.

As for our conjecture, the remainder of the discussion will be restricted
to the \newspace\ with $p=3$, 
\ie spacetime dimension $d=5$.
In this case, the metric on a constant time slice in \threebrane\ is
\eqn\slice{
ds^2={r^2\over \l^2}\left[\left(1-{r_0^4\over r^4}\right) d\tau^2 
+(dx^1)^2+(dx^2)^2\right]+\left(1-{r_0^4\over r^4}\right)^{-1}{\l^2\over r^2}
{dr^2}\ . }
In the following, we will actually refer all indices to the obvious orthonormal
frame. Now the curvature of this spatial slice is given by
\eqn\newcurv{\eqalign{
\upp \bR_{\tau r\tau r}&=-{1\over \l^2}(1-3y)\qquad
\upp\bR_{1212}=-{1\over \l^2}(1-y)\cr
\upp\bR_{\tau 1\tau 1}&=-{1\over \l^2}(1+y)=\upp\bR_{\tau 2\tau 2}=
\upp\bR_{r 1r 1}=\upp\bR_{r 2r 2} \cr}}
where $y=r_0^4/r^4$. Now the potential term in eq. \hamb\ becomes
\eqn\newpot{\eqalign{
U&={1\over2}(\,\upp{\bR}^{ijkl}h_{il}h_{jk}-
\upp{\bR}^{ij}h_{ik}h_j{}^k\,)\cr
&={1\over \l^2}\left[(2-y)((h_{\tau 1})^2+(h_{\tau 2})^2+(h_{r1})^2
+(h_{r2})^2)\right.\cr
&\qquad\qquad\left.+2(1+y)(h_{\tau r})^2+2(h_{12})^2
+U_{\rm diag}(h_{\tau\tau},h_{rr},h_{11},h_{22}) 
\right]\cr} }
Given that $0\le y\le1$, we see that the potential ensures the
stability of  all fluctuations
in the off-diagonal components of the metric. Now in evaluating the potential
for the diagonal fluctuations, we first impose the traceless condition
of eq. \constrain\ with $h_{rr}=-h_{\tau\tau}$--$h_{11}$--$h_{22}$. Then
defining $V^a=(h_{\tau\tau},h_{11},h_{22})$, the remaining potential
terms may be written as
\eqn\newpotd{
U_{\rm diag}={1\over2\l^2}V^aU_{ab}V^b }
where
\eqn\matu{
U_{ab}=2\pmatrix{2+2y&1+y &1+y\cr
                 1+y&2-y &1-2y\cr
                 1+y&1-2y&2-y \cr} }
It is straightforward to determine the eigenvalues of this matrix to
be
\eqn\eigenv{
\la_0=2(1+y),\qquad\la_\pm=5-y\pm(9+6y+33y^2)^{1/2} }
One easily shows that $\la_0$ and $\la_+$ are  positive in the
range of interest, \ie $0\le y\le1$, and hence the corresponding
eigenvectors correspond to manifestly stable metric fluctuations.
The most interesting case is that of $\la_-$ for which one finds
$\la_->0$ for $0\le y<{1\over2}$ but $\la_-<0$ for ${1\over2}<y\le1$.
Hence the potential energy density for this eigenvector
\eqn\eigenvect{
V_-^a=(v_-,1,1)\qquad{\rm with}\ \ \ v_-={-1+5y-(9+6y+33y^2)^{1/2}\over
2(1+y)} }
 becomes
negative in a small region near the center of the space, \ie $r^4<
2r_0^4$.

Thus metric fluctuations with a form where the 
$h_{ij}$ are
dominated by this eigenmode and only have support in this small region
near $r=r_0$ seem to have the potential to be unstable, \ie lower the
energy of the background solution.
 However for such fluctuations, there is a competition between
the manifestly positive gradient contributions and the potential
terms in the energy density \hamb. We argue below that the former terms
dominate and hence these fluctuations are also stable.

Imagine that we are considering a metric fluctuation which we might
characterize as $V^a=A(r)\,V_-^a$ where $A(r)$ is a profile, which we assume
takes its maximum at $r=r_0$, monotonically decreases, in order
to minimize the gradient energy, and vanishes outside $r=2^{1/4}r_0$.
The potential
energy density \newpotd\  becomes $U={A^2\over2\l^2}\la_-(2+v_-^2)$.
With the assumption that the profile takes its maximum at $r=r_0$,
the minimum of the potential energy density is 
\eqn\umin{
U_{\rm min}=U(r=r_0)=-8(2\sqrt{3}-3){A(r_0)^2\over \l^2}
\simeq -3.713{A(r_0)^2\over \l^2} }
While the complete expression for $U$ for this fluctuation
 has a complicated analytic form,
for the purpose of the reader's intuition we note that to within an
accuracy of a few percent one may approximate this expression in the
region of interest, \ie ${1\over2}<y\le1$, by the simple expression
$U=U_{\rm min}\,(2y-1)A(r)^2/A(r_0)^2$. So if one imagined that the profile
was constant, the potential energy density would decrease linearly
as a function of $y$ reaching zero at $y=1/2$.

We estimate the gradient energy as follows
\eqn\kinetic{
T={1\over4}(\bD h)^2\simeq {1\over4}(2+v_-^2+(2+v_-)^2)(\bD A)^2 }
where we further estimate the gradient of the profile by its maximum
$A(r_0)$ divided by the proper distance between $y=1$ and $y=1/2$,
\ie between $r=r_0$ and $r=2^{1/4}r_0$. The latter distance turns out
to be $\l\,\log(\sqrt{2}+1)/2\simeq.441\,\l$. In the region of interest,
$v_-$ varies from --1 at $y=1/2$ to $1-\sqrt{3}\simeq-.73$ at $y=1$
so we will simply fix it to $v_-=-1$ in our estimate of $T$.
Hence we arrive at the following estimate
\eqn\kinetid{
T_{\rm average}\simeq 5.14 {A(r_0)^2\over \l^2} }
Comparing eqs. \umin\ and \kinetid, we see that this average gradient
energy already exceeds the minimum value of the potential energy.
Hence it must be that these potentially unstable
metric fluctuations in fact have a positive total energy.
While our estimate of the gradient contribution may seem crude, a
more detailed examination shows that in fact it greatly underestimates
the energy. Properly evaluating the covariant derivatives in eq. \kinetic\ 
accounting for the tensor properties of the fluctuations
adds more positive terms to this expression, which are roughly the
same order of magnitude as those considered, \ie $A^2/\l^2$. Further we have
not accounted for the gauge fixing constraint \gauge\ in our calculations.
This constraint fixes the form of the profile for the fluctuation
considered above through $\bD^ih_{ir}=0$, which is the only nontrivial
component. One finds that the profile must decay more slowly than estimated
above,
but that it cannot vanish at $y=1/2$. Rather it has infinite support, vanishing
as $1/r^4$ in the asymptotic region. While this decreases the local gradient
energy density, it also adds a positive potential energy density in the
region $y<1/2$. Hence the final conclusion that these fluctuations are
stable remains correct in a detailed analysis.

\newsec{Generalizations}

Although we have focussed on the case $p=3$ above, one can extend the
conjecture to other dimensions, including the $p=5$ case which is
directly related to four dimensional nonsupersymmetric gauge theories \witb.
We believe the solutions \aneext\ for all $p$ are
perturbatively stable, although a
detailed analysis of the fluctuations has not yet been carried out.

Consider the following modification \newmet\ of the metric \sadsall:
\eqn\sadsal{
ds^2 = - \left[{r^2\over \l^2} -1 -\left({r_0\over r}\right)^{p-1}\right]
dt^2 + \left[{r^2\over \l^2} -1 -\left({r_0\over r}\right)^{p-1}\right]^{-1}
{ dr^2} + r^2 d\sigma^2_p}
where 
\eqn\omegak{
d\sigma^2_p = (1+\rho^2) dz^2 + {d\rho^2\over 1+\rho^2} + \rho^2 d\Omega_{p-2}}
is the metric on a $p$-dimensional unit hyperboloid. 
This metric is also a solution of Einstein's equation with negative 
cosmological constant, and is asymptotically AdS. 
It is unusual since the black hole metric \sadsal\
(without any analytic continuation) can have negative energy \newmet. 
Indeed, the energy
is still given by \energee\ (with $\Omega_p$ denoting the area of the unit
hyperbolic
space\foot{One can either compactify this space so that its area is finite, or
work with the energy density.}) but now one can let the parameter
$r_0^{p-1}$ be negative and still have a 
horizon. There is, however, a minimum energy possible for the black hole.
For example, in the case $p=3$, 
this occurs when $r_0^2 = -\l^2/4$, corresponding
to an energy
\eqn\negbh{ E=-{3\l^2 \Omega_3\over 64 \pi G_5}}
The Hawking temperature of these minimum energy black holes is zero. 
Are they the minimum energy configurations with these boundary conditions?

One can double analytically continue this metric, $t \rightarrow i\tau$
and $z\rightarrow it$. As usual, $\tau$ must be periodically identified
with period
\eqn\newperid{
\beta = {4\pi \l^2 r_+\over (p+1)r_+^2 - (p-1)\l^2}}
so a constant time surface asymptotically approaches the product of hyperbolic
space and a circle.
The energy is again given by \newenergy\ and hence is negative for positive
values of $r_0^{p-1}$. For large $r_0^{p-1}$, the energy is the same as the
metric \highads,
since the size of the circle is much smaller than the scale
of the curvature on the orthogonal space. 

One might conjecture that the metrics \sadsal\ represent the minimum energy
configurations for these boundary conditions.
Since these solutions are static, they are extrema of the energy. It is likely
that they are also local minima of the energy. In fact the calculations
in section 4 show that this is the case for $p=3$. This is due to the
remarkable fact that the components of the curvature of \sadsal\ (in an
orthonormal frame) are identical to \sadsall!\foot{This
provides a counterexample
to the popular idea that if two metrics have the same curvature, they are
locally isometric. The covariant derivatives of the curvature are different,
showing that the metrics are indeed inequivalent.
We thank S. Ross for discussions on this point.} If one analytically 
continues in $t$ and $z$ and restricts to a constant time surface, the
curvatures are still the same. Since the potential term in the
quadratic fluctuations depends only on  the curvature, it will again be 
positive. 

{}From the CFT viewpoint, the above metrics should describe the CFT on
a product of $S^1$ and a hyperbolic space. Since the scalars couple
to the curvature, a negative curvature space would seem to lead to an
instability.\foot{We thank J. Maldacena for pointing this out.}
Thus, in this case, the apparent stability of the supergravity
solution seems in contrast to the expected CFT result. We do not yet
understand the resolution of this puzzle.

\newsec{Discussion}

We have shown that the AdS soliton \aneext\ has lower energy than AdS
itself. Rather than producing
a contradiction with the recently conjectured AdS/CFT correspondence,
these results find close agreement with the negative Casimir energy of
nonsupersymmetric field theory on $S^1\times R^{p-1}$. 

Since the AdS soliton has extended translational symmetry,
one can define not just the total energy, but a full boundary
stress-energy tensor. One finds that the agreement in the energy
densities, discussed here, extends to agreement of all of the components
of the stress tensor \prep. In the example of the gauge theory on
$S^1\times R^2$, one finds that the factor of $3/4$ relating the
energy densities calculated with supergravity and in the weakly
coupled gauge theory is, in fact, an overall factor relating the
full stress tensors calculated in these two regimes.

It is examining the full stress-energy tensor which motivated
in part our choice of physical boundary conditions \boundary\ 
in the positive energy conjecture and in our discussions
in the earlier sections. For example, if one retains the $r_1/r$
term in our first negative energy example \rnads, one might expect
that the energy would be divergent. One finds though that there
is a precise cancellation in the calculation so that the final
result remains exactly the same as in \maschw, which was derived for
$r_1=0$. This is also true for the translationally invariant solutions.
However, if one considers the full boundary stress tensor,
a nonvanishing $r_1$ produces divergences in the spatial components
of the stress-energy. One should expect that a less symmetric choice
of the initial data surface would yield an energy density which mixes
the various components of the latter tensor, and hence diverges.
Our physical boundary conditions \boundary, which rule out including $r_1\ne0$,
ensure that the energy density will remain finite for any choice
of time slicing. 

A precise comparison of the supergravity and gauge theory energies was
only attempted for $p=3$ because this is the case in which the AdS/CFT
correspondence is best understood. For this dimension, we only considered
the \newspace\ solution, which corresponds to the gauge theory on
$S^1\times R^2$. However, one might also consider the initial data
\rnadsn, which
would correspond to the gauge theory on $S^1\times S^2$. We calculated
the supergravity energy \maschw, and need only translate it to a
gauge theory expression, as in going between \massane\ and \finaleng.
The final result is that \maschw\ yields
\eqn\spheric{
\rho_{\rm sugra} =-{\pi^2 \over 8}
{N^2\over\beta^4}\,F\left(\beta^2/\l^2\right) }
where the function satisfies $F(0)$=1. Hence to leading order
for small $\beta$, the energy density is precisely the same as
for $S^1\times R^2$. This is not surprising as this result is valid
in the limit
where the radius of the sphere  is much bigger that the period
of the circle $\l \gg \beta$,
and so the $S^2$ factor looks essentially like a flat $R^2$
 on the latter scale. One should note that this energy is measured
 relative to the standard AdS background with periodic identifications,
which naturally includes the curved $S^2$ factor in its asymptotic
geometry. Thus this negative energy does not include the positive
contribution which might be expected to appear as the Casimir energy
of the two-sphere. It is not difficult to calculate the complete
 function 
\eqn\function{
F(x)={1\over16}\left(1+\sqrt{1-{2x\over\pi^2}}\right)^2
\left[\left(1+\sqrt{1-{2x\over\pi^2}}\right)^2+{4x\over\pi^2}\right]\ .}
Taylor expanding $F$ for small $x$ would yield higher order corrections
to the energy density.
It would be interesting to see to what extent the coefficients of
the higher order terms are reproduced in the weakly coupled regime of
the gauge theory. This function is only defined for $x\le \pi^2/2$ since
these are the allowed values of $\beta^2/\l^2$ from \period. (The solutions
\rnadsn\ only exist if the circle is small enough.) This suggests that
there might be a jump in the ground state energy of the strongly 
coupled gauge theory on $S^2\times S^1$, analogous to the
Gross-Witten-like phase transition \grosswit\
discussed in \wita.

The agreement of the negative energy of the AdS soliton
with the expected negative
Casimir energy appears to support the AdS/CFT correspondence even
in the nonsupersymmetric case. But this is true only if the AdS soliton
is the lowest energy solution with the given boundary conditions. This
gives rise to a new type of positive energy conjecture. 
Although this conjecture seems  unlikely to be true from a 
purely mathematical standpoint,  we have presented evidence to support it, 
by showing that the AdS soliton is indeed a local
minima of the energy (for the  case $p=3$). It is natural to extend
this conjecture to other dimensions and other asymptotic boundary conditions
as we discussed in  the previous section.

Our analysis of the perturbative stability of the \newspace\ 
with respect to metric fluctuations is closely related to the recent
calculations of glueball masses in large N QCD \glue. General arguments
have been given that massless supergravity fields propagating on the
\newspace\ background will have a discrete spectrum \witb. Further,
by the AdS/CFT correspondence, these fluctuations should correspond to
various glueball states in the large N gauge theory \refs{\witb,\grossooguri}.
In the context of our first form of the conjecture,
the given calculations of glueball spectra \glue\ verify that
the \newspace\ is stable against fluctuations of many of the
bosonic supergravity fields, \eg the dilaton, and Neveu-Schwarz and
Ramond-Ramond antisymmetric tensors. 
While our perturbative calculations have not produced a precise spectrum,
they do verify that a positive mass gap exists for the metric fluctuations,
\ie the spin-two glueballs.
It is interesting that amongst the metric excitations, our calculations
indicate that the lowest energy state (\ie the mode described as
potentially unstable) must in fact contain a scalar contribution
(with respect to $R^2$) which should actually decouple
as the ultraviolet regulator is removed.

Lest the readers imagine that the AdS/CFT correspondence guarantees
the local stability of all (static) supergravity solutions, we remind
them that this is not the case. Typically there
are many unstable stationary points
in the scalar potential of the gauged supergravity theory -- see,
\eg \huwa. For these stationary points,
there will be an AdS background, but the cosmological constant will
have a value such that supersymmetry is completely broken. In the
supergravity analysis, the instability arises because some of the
scalars have curvature couplings which exceed the Breitenlohner-Freedman
bound \stable, \ie scalar fluctuations around the stationary point
have masses which are (too) negative. It would be interesting to
understand what the corresponding physics in the gauge theory
is.

\vskip 1cm
{\bf Acknowledgments}

We thank G. Gibbons and
the participants of the String Duality program at the Institute
for Theoretical Physics, Santa Barbara for helpful discussions.
GTH was supported in part by NSF Grant PHY95-07065.
RCM was supported in part by NSERC of Canada and Fonds FCAR du Qu\'ebec.
Both GTH and RCM were supported by NSF Grant PHY94-07194
while at the ITP.

\listrefs
\end